\documentclass{elsart}
\usepackage{graphicx}
\usepackage{bm}
\usepackage{amsmath}
\usepackage[centerlast]{subfigure}
\begin{document}
\begin{frontmatter}

\title{Large-scale effects on meso-scale modeling for scalar
transport}
\author{M. Cencini$^{1,2}$,
        A. Mazzino$^{3}$, 
        S.~Musacchio$^{2,4}$ and 
        A.~Vulpiani$^{2,4,5}$}

\address{$^{1}$ ISC-CNR, Via dei Taurini 19, I-00185 Roma, Italy }
\address{$^{2}$ INFM-SMC c/o Dipartimento di Fisica, Universit\`a ``La
                Sapienza'',\\ P.zzle Aldo Moro 2, I-00185 Rome,
                Italy.}  \address{$^{3}$ Dipartimento di Fisica,
                Universit\`a di Genova and INFN Sezione di Genova, Via
                Dodecanneso 33, I--16146 Genova, Italy.}
                \address{$^{4}$ Dipartimento di Fisica and INFM UdR
                Roma 1, Universit\`a ``La Sapienza'',\\ P.zzle A. Moro
                2, I-00185 Roma, Italy.}  \address{$^{5}$ INFN -
                Sezione di Roma ``La Sapienza''.}

\vskip 1.5cm

\begin{abstract}
\noindent
The transport of scalar quantities passively advected by velocity
fields with a small-scale component can be modeled at meso-scale level
by means of an effective drift and an effective diffusivity, which can
be determined by means of multiple-scale techniques.  We show that the
presence of a weak large-scale flow induces interesting effects on the
meso-scale scalar transport.  In particular, it gives rise to
non-isotropic and non-homogeneous corrections to the
meso-scale drift and diffusivity.  We discuss an approximation that
allows us to retain the second-order effects caused by the large-scale
flow.  This provides a rather accurate meso-scale modeling for both
asymptotic and pre-asymptotic scalar transport properties.  Numerical
simulations in model flows are used to illustrate the importance of
such large-scale effects.
\end{abstract}

\end{frontmatter}
\section{Introduction}
\label{sec:1}

Diverse scientific disciplines cope with systems characterized by
interactions across a large range of physically significant length
scales. This is for instance the case of earthquake dynamics (see,
e.g., Refs.~\cite{MRAK02,T97}), where the important spatial
scales range from the small scale ($1~cm$ to $1~m$), associated
with friction, to the tectonic plate boundary scale ($10^3 - 10^4~km$).
The range of temporal scales goes from seconds (during dynamic
ruptures) to $10^3 - 10^4$ years (repeat times for earthquakes) to
$10^7 - 10^8$ years (evolution of plate boundaries).

Many active scales are also encountered in biological and soft matter
sciences. For instance, to understand the physics of a cell membrane,
one starts from Angstrom-sized atoms with motions in the femtosecond
range, to go to the whole cell, whose diameters can be tens of
micrometers and lifetimes of the order of days.  All these scales can
neither be probed by a single experimental technique, nor numerically
resolved in a single simulation (see e.g.~Ref.\cite{MFPKV04}).

A similar situation can be found in the framework of atmospheric or
ocean sciences. Focusing on tracer dispersion, advection by
geophysical flows results in the rapid generation of fine-scale tracer
structures. This is true even though, e.g. in the atmosphere, the wind
field is typically dominated by large scale features, such as synoptic
scale weather systems. Coarse-grained representations of atmospheric
winds can indeed generate fine-scale tracer filaments, through the chaotic
advection process (see e.g.~Ref.~\cite{O89}).

A common and long-standing problem linking all these scientific
disciplines is to build a coarse-grained description starting from the
microscopic model.  The challenge is thus to reduce the typically huge
number of degrees of freedom by modeling the fastest and smallest
scales. In this way one may construct a computationally tractable
effective equation, which involves only the scales one is interest in
\cite{LES,MK99}.

Our aim here is to shed some light on this very general problem in the
specific framework of tracers dispersion. In this context, the scales
of interest are generally larger than the diffusive length-scale. The
problem of developing coarse-grained models, closed on the large
scales, is also known as ``parameterization problem'' \cite{S88}. 
For general advecting velocity fields, the
parameterization problem is not tractable by means of systematic
approaches able to deduce the form of the closure starting from the
original, full-scale, equations. 

An important exception occurs in the presence of scale separation
between the advecting velocity and the tracer i.e., when the former is
concentrated at small scales and one looks at the tracer dynamics on
scales much larger than those of the velocity.  In this limit the goal
is to derive the expression of the asymptotic diffusion coefficient
renormalized by the presence of the small scale velocity field.  This
can be accomplished exploiting asymptotic methods (see, e.g.,
\cite{MK99,BLP78,Piretal,BGW89,BCVV95,VA97,PK02,M03} among the
others).  However, in many physical circumstances one has that the
velocity field may be though as a small-scale advecting velocity field
(at scale $\ell$) superimposed to a large-scale, slowly varying
component (at scale $L\gg \ell$). A physical example of this situation
is represented by the dispersion of a pollutant that, released in
the atmospheric boundary layer (having characteristic time scales of
the order of minutes), is observed at synoptic scales (i.e. over weeks
or more).  Scale separation is guaranteed by the fact that meso-scale
structures (in the range between weeks and minutes) turns out to be
much less energetic than the other active scales (see e.g.~
Ref.~\cite{MY75}).

In this case two levels of coarse-grained description are of interest.
As in the previous case one would like to understand the asymptotic
diffusive regime at scales much larger than $L$. On the other hand,
one is also interested in deriving meso-scale models for the
(pre-asymptotic) transport of a tracer in the range of scales 
between $\ell$ and $L$. The latter issue has been recently addressed in
Refs.~\cite{M97,MMV05}.  In particular, explicit expressions for the
effective (meso-scale) diffusivity and velocity have been obtained in
the limit of strong large scale flows \cite{MMV05}.  This
investigation pointed out the important fact that the (meso-scale)
diffusivity does depend on the large-scale advecting velocity, as well
as (of course) on the small scales.

Our main aim here is to quantitatively understand whether or not the
opposite limit (i.e.~a large-scale advective component which is weaker
than the small-scale advection) leads to explicit expressions for the
effective parameters which functionally depend on the large-scale
advecting velocity.

As we shall see, a class of eddy-diffusivity fields will emerge from a
perturbative approach through which both the asymptotic and the
pre-asymptotic transport properties can be successfully described when
the effects of the large scales are properly taken into account.  In
particular, by means of numerical simulations of model flows we show
that the meso-scale transport model is able not only to recover the
asymptotic properties but also to predict the pre-asymptotic regime.
The latter issue is conceptually connected with the predictability
problem of the second kind \cite{L96,BCFV02}, where the goal is to
predict the evolution of the system through a model in which not all
the degrees of freedom are resolved.

For the sake of simplicity our analysis will be restricted to
two-dimensional flows. It should be noted that working in
two-dimensions is still relevant to many applications, e.g., to
investigate the time-varying transport and mixing properties of
isentropic flows in the atmosphere (see e.g. Ref.~\cite{HS00}) and in
the ocean in connection with horizontal geostrophic eddies (see
e.g. Ref.~\cite{MSJH05}).

The material is organized as follows.  In Sec.~\ref{sec:2} the general
framework for the multiple-scale analysis for a passive tracer in the
presence of a slowly varying advective velocity superimposed to a
small-scale component is described, and the limit of weak large-scale
advection is explicitly considered. In Sec.~\ref{sec:flows} eddy
diffusivities are computed in two commonly considered model flows.  In
Sec.~\ref{sec:results}, the meso-scale model is compared with direct
numerical simulations of the original transport problem both for the
asymptotic and pre-asymptotic properties.  Conclusions and
perspectives are reserved to Sec.~\ref{sec:conclu}.  The Appendix
presents some technical material.

\section{Multiple-scale analysis}
\label{sec:2}
We consider the evolution of a passive scalar field, $\theta({\bm x},t)$,
in an incompressible velocity field $\bm{v}$:
\begin{equation}
\partial_t\theta(\bm{x},t)+ 
(\bm{v} \cdot \bm{\partial})\theta(\bm{x},t)=
D_0\partial_i\partial_i \theta (\bm{x},t) \;\;\; ,
\label{eq:1}
\end{equation}
where $D_0$ is the molecular diffusivity. Following
Refs.~\cite{M97,MMV05}, we focus on situations where $\bm{v}$ can be
thought as the superposition of  a ``large-scale'' velocity field
$\bm{U}(\bm{x},t)$ and a ``small-scale'' component $\bm{u}(\bm{x},t)$
which are assumed to vary on length-scales of order $O(L)$ and
$O(\ell)$, respectively.  Scale separation between the two fields is
measured by the small parameter $\epsilon = \ell/L$. In the limit
$\epsilon \to 0$, multiple-scale analysis provides a description for
modeling the dynamics of the scalar field at meso-scale, i.e. at scales
larger than $\ell$ and of the same order of $L$, in which the
dynamical effects of the smallest scales appear via a renormalized
(enhanced) diffusivity \cite{M97,MMV05}.

Following the multiple-scale approach we introduce a set of {\em slow
variables} ${\bm X}=\epsilon {\bm x}$, $T={\epsilon}^{2} t$ and
$\tau=\epsilon t$ in addition to the {\em fast variables} $({\bm
x},t)$.  The scaling of the times $T$ and $\tau$ are suggested by
physical reasons: we are searching for diffusive behavior on large
time scales of $O(\epsilon^{-2})$ taking into account the effects
 played by the advection contribution occurring on time scales of
$O(\epsilon^{-1})$. According to the prescription of the method, the
two sets of variables are treated as independent.  It then follows
that
\begin{equation}
\label{derivate}
\partial_i\mapsto\partial_i+\epsilon\nabla_i\,,\qquad
\partial_t\mapsto\partial_t+\epsilon\partial_{\tau} +\epsilon^2\partial_T,
\end{equation}
\begin{equation}
\label{campi}
\bm{u}\mapsto\bm{u}(\bm{x},t)\,,\qquad \bm{U}\mapsto\bm{U}(\bm{X},T),
\end{equation}
where $\partial$ and $\nabla$ denote the derivatives with respect to
fast and slow space variables, respectively. 

The passive scalar field is expanded as a perturbative series 
\begin{equation}
\label{espansione_campo}
\theta(\bm{x},t;\bm{X},T;\tau) = \theta^{(0)} + 
\epsilon \theta^{(1)} + \epsilon^2 \theta^{(2)} + O(\epsilon^3) \;.
\end{equation}
By inserting Eqs.~(\ref{derivate}) and~(\ref{espansione_campo}) into
Eq.~(\ref{eq:1}) and equating terms with equal powers in $\epsilon$
one obtains a hierarchy of equations. By imposing the solvability
conditions on the first two orders in $\epsilon$ one derives
a Fokker-Planck equation for the ``large-scale'' scalar field defined
as $\theta_L \equiv \langle \theta^{(0)} \rangle + \epsilon \langle
\theta^{(1)} \rangle$ (see Ref.~\cite{M97} for a detailed derivation).

The eddy-diffusivity tensor of this ``coarse-grained'' Fokker-Planck
equation is given by
\begin{equation}
D_{ij}(\bm{X},T)=\delta_{ij}D_0- \langle u_{i} \chi_{j} \rangle
\label{eq:3} 
\end{equation}
where brackets indicates spatial and temporal averages
over the fast variables $\bm x$ and $t$ and $\bm{\chi}(\bm{ x},t ;\bm{ X},T)$ 
is an auxiliary field with vanishing average over the periodicities that  
obeys to the following dynamics 
\begin{equation}
\partial_t { \chi_j}+\left[(\bm{ u}+ \bm{ U})\cdot\bm{ \partial}\right]
{ \chi}_j - D_0\, \partial^2{ \chi}_j = -{u}_j \;.
\label{eq:4}
\end{equation}
The meso-scale transport equation for the ``large-scale'' scalar field
$\theta_L$ reads
\begin{equation}
\partial_{t} \theta_L +(\bm{U}\cdot\bm{\partial}) \theta_L =\partial_{i}
\left( D_{ij} \partial_j \theta_L \right)
\label{eq:2}
\end{equation}
where, for the sake of notation simplicity, the usual variables $t$ and
${\bm x}$ have been restored.  
We remind that Eq.~(\ref{eq:2}) is the Fokker-Plank equation corresponding 
(in the  Ito convention) to the Lagrangian description
\begin{equation}
\label{eq:ito}
\frac{d x_i}{dt}=  U^{E}_i+B_{ij}\eta_j
\end{equation}
where $U^{E}_i=U_i + \partial_j D_{ij}$ is the effective meso-scale velocity, 
$B_{ik}B_{jk}= D_{ij} + D_{ji}$ and $\eta_i$'s are zero-mean
Gaussian variables with $\langle
\eta_i(t)\eta_j(t')\rangle=\delta_{ij}\delta(t-t')$.

It is worth noticing that the multiple-scale approach reduces the
calculation of eddy diffusivities and meso-scale velocities to the
solution of the auxiliary equation (\ref{eq:4}). In generic flows,
when $\bm{U}$ is not a constant mean flow but depends on $\bm{X}$ and
$T$, the equation must be solved for each value of $\bm{U}$. This can
be rather demanding in terms of computer resources if numerical
methods are required to solve such an equation. Therefore, except for a
few cases, in which analytic solutions of Eq.~(\ref{eq:4}) are
available (e.g. in the case of orthogonal shears \cite{M97}, this
approach does not provide a practical tool for evaluating the
eddy-diffusivity.

\subsection{Weak large-scale flow asymptotics}
The natural way to overcome the problem of finding solutions of
Eq.~(\ref{eq:4}) for arbitrary values of the large-scale flow is to
try to find explicit, even if approximate, expressions for the
eddy-diffusivities.  In this section we show that this program can be
fulfilled in a perturbative limit. To be more specific, here we
consider situations in which the intensity of the large-scale flow is
weak compared to the small-scale one (the opposite limit was
considered in Ref.~\cite{MMV05}).  In this case it is possible to find
the solution of Eq.~(\ref{eq:4}) as a perturbative expansion in the
small parameter $\varepsilon = U/u$:
\begin{equation}
{\bm \chi}({\bm x},t;{\bm X},T) =  
{\bm \chi}^{(0)} + \varepsilon {\bm \chi}^{(1)} + 
\varepsilon^2 {\bm \chi}^{(2)} +\dots\;.  
\label{eq:3.2}
\end{equation}
Consistently, from Eq.~(\ref{eq:3}) the eddy diffusivity can be written as  
\begin{equation}
D_{ij}(\bm X,T) = D_0 \delta_{ij} - \langle u_i \chi^{(0)}_j \rangle -
\varepsilon \langle u_i \chi^{(1)}_j \rangle - \varepsilon^2 \langle
u_i \chi^{(2)}_j \rangle + O(\varepsilon^3)\,.
\label{eq:13}
\end{equation}

Defining the $O(1)$ field, ${\bm U'} = {\bm U} / \varepsilon$, plugging
Eq.~(\ref{eq:3.2}) into Eq.~(\ref{eq:4}) and equating terms with equal
powers in $\varepsilon$, one obtains the following hierarchy of
equations:
\begin{eqnarray}
\partial_t {\bm \chi}^{(0)} + 
({\bm u} \cdot {\bm \partial}) {\bm \chi}^{(0)}
- D_0 \partial^2 {\bm \chi}^{(0)} & = & 
-{\bm u} \;, 
\\
\label{eq:5} 
\partial_t {\bm \chi}^{(1)} + 
({\bm u} \cdot {\bm \partial}) {\bm \chi}^{(1)}
- D_0 \partial^2 {\bm \chi}^{(1)} & = &  
-({\bm U'} \cdot  {\bm \partial}) {\bm \chi}^{(0)}  \;, 
\\
\label{eq:6}
\cdots \cdots \phantom{-----------} & & \nonumber \\
\partial_t {\bm \chi}^{(n)} + 
({\bm u} \cdot {\bm \partial}) {\bm \chi}^{(n)}
- D_0 \partial^2 {\bm \chi}^{(n)} & = &  
-({\bm U'} \cdot {\bm \partial}) {\bm \chi}^{(n-1)} \;.
\label{eq:7}
\end{eqnarray}
The zeroth-order solution does not depend on the large-scale variables, 
and can be formally written as:
\begin{equation}
\chi_i^{(0)}({\bm x},t) = 
- \int G({\bm x}-{\bm x'},t-t') u_i({\bm x'},t')\;d{\bm x'}\;dt' \;,
\label{eq:8}
\end{equation}
where $G({\bm x},t)$ is the Green function associated to the
linear differential operator of the Fokker-Planck equation:
\begin{equation}
{\mathcal L}_{FP} G(\bm{x},t) = \left\{ \partial_t  + 
({\bm u} \cdot {\bm \partial}) 
- D_0 \partial^2  \right\} G(\bm{x},t)  = \delta(\bm{x},t) \;.
\label{eq:9}
\end{equation}
In the same way the $n$-th order solution can be written as
\begin{equation}
\chi_i^{(n)}({\bm x},t;{\bm X},T) = - \sum_{j} U'_j ({\bm X},T) 
\int G({\bm x}-{\bm x'},t-t') \partial_j \chi_i^{(n-1)} ({\bm x'},t';\bm X,T)
\;d{\bm x'}\;dt' \;.
\label{eq:10}
\end{equation}
Let us notice that all the odd terms of the expansion in
Eq.~(\ref{eq:13}) must vanish by symmetry: due to the recursive nature
of Eq.~(\ref{eq:10}) they correspond to polynomial correction of odd
order in ${\bm U}$ to the eddy diffusivity, which would depend on the
sign of ${\bm U}$.

Plugging Eqs.~(\ref{eq:8}) and~(\ref{eq:10}) into Eq.~(\ref{eq:13}) we
can finally write the following polynomial expansion in ${\bm U}$ for
the eddy diffusivity:
\begin{equation}
D_{ij}(\bm X,T) = 
D^S_{ij} + \sum_{lm} U_l U_m \Gamma^{lm}_{ij} + O(\varepsilon^4) 
\label{eq:11}
\end{equation}
where we denoted with $D^S_{ij}$ the eddy
diffusivity originating from the small-scale velocity field when the
large-scale one is neglected, i.e. :
\begin{equation}
D^S_{ij} = D_0 \delta_{ij} - \langle u_i \chi^{(0)}_j \rangle \,.
\label{eq:14}
\end{equation}
It is important to note that the coefficients 
\begin{equation}
\Gamma^{lm}_{ij} = 
\langle u_i G \ast (\partial_l (G \ast (\partial_m (G \ast u_j)))) \rangle 
\label{eq:coeffgamma}
\end{equation}
(where $\ast$ denotes the convolution integral) are determined only by
the small-scale flow characteristics. We remark
that this approach is relevant also in those cases in which direct
(analytical or numerical) computation of these coefficients is not
possible, since they can be guessed on an empirical ground.

It is worth recalling that the expression (\ref{eq:11}) for the eddy
diffusivity is a perturbative expansion in two different small
parameters: $\epsilon = \ell /L$ (which measures the scale separation
between the two component of the flow), and $\varepsilon = U/u$ (which
measures the ratio of their intensity).  The first expansion is the
basis of the multiple-scale approach.  The second allows to
disentangle the dependence on the large-scale flow in the eddy
diffusivity, since it provides recursive expressions for the auxiliary
fields $\chi^{(n)}$ as polynomial expansion in ${\bm U}$.

In order to illustrate how Eq.~(\ref{eq:11}) can be used to retain the
effects induced by the large-scale flow on the eddy diffusivity in
meso-scale modeling, let us concentrate on the first non-vanishing
correction, i.e. the second-order one in $\varepsilon = u /U$: $\delta
D^{(2)}_{ij}(\bm X,T) = - \varepsilon^2\langle u_i \chi^{(2)}_j
\rangle $.  For the sake of simplicity, we consider the case of a two
dimensional flow in the case with ${\bm u}$ is statistically isotropic.  
Without loss of generality one can
choose the $x'_1$ axis parallel to ${\bm U}$. In this coordinate system 
\begin{equation}
\delta D^{(2)}_{i'j'}(\bm X,T) = 
U^2(\bm X,T) \langle u_{i'} G \ast (\partial_{1'} 
( G \ast ( \partial_{1'} (G \ast u_{j'})))) \rangle 
\label{eq:17} 
 \end{equation}
The off-diagonal terms $\delta D^{(2)}_{1'2'}(\bm X,T)$ and $\delta
D^{(2)}_{2'1'}(\bm X,T)$ vanish by isotropy. The correction to
$D^S_{ij}$ is then diagonal in the reference frame with axis $X'_1
\parallel {\bm U}$ and $X'_2 \perp {\bm U}$.  Finally we can write
\begin{equation}
D_{ij}(\bm X,T) = 
D^S_{ij} + \delta D^{(2)}_{ij}(\bm X,T)+O(\varepsilon^4) \;,
\label{eq:19}
\end{equation} 
where
\begin{equation}
\delta D^{(2)}_{ij}(\bm X,T) = U^2(\bm X,T) R(\phi) 
\left( \begin{array}{cc}   
\alpha & 0 \\
0 & \beta 
\end{array} \right) 
 R^T(\phi) \;,
\label{eq:20}
\end{equation}
with
\begin{equation}
R(\phi) = \frac{1}{U} 
\left( \begin{array}{cc}   
U_1 & - U_2 \\
U_2 & U_1 
\end{array} \right)
\label{eq:21}
\end{equation}
being the rotation matrix and $\phi$ the angle between ${\bm U}$ and
$x_1$.  Therefore the effects induced by the large-scale flow on the
eddy diffusivity, up to the second order, can be obtained in terms of
two parameters only. These  are determined solely by the small-scale
features:
\begin{eqnarray}
\alpha &=& \langle u_\parallel G \ast (\partial_\parallel 
( G \ast ( \partial_\parallel (G \ast u_\parallel))))\rangle \;,
\label{eq:22} \\
\beta &=& \langle u_\perp G \ast (\partial_\parallel 
( G \ast ( \partial_\parallel (G \ast u_\perp)))) \rangle \;.
\label{eq:23}
\end{eqnarray}
In the general three-dimensional case the transverse correction $\beta$ 
pertains to the plane perpendicular to the direction of the
large-scale flow. 

\section{Computation of the eddy diffusivity in model flows}
\label{sec:flows}
Let us now discuss the approximation previously obtained in two model
flows. As representative examples of two broad classes of realistic
instances we focus here on two large-scale flows: 
case $(a)$: a steady shear, the Kolmogorov flow~\cite{MS61}  
\begin{equation}
{\bm U} = (U \sin(Ky),0)  
\label{eq:3.26}
\end{equation}
and case $(b)$: a large-scale cellular flow~\cite{MK99,sg88,BCVV95} 
\begin{equation}
{\bm U} = (U \sin(Kx) \cos(Ky), -  U \cos(Kx) \sin(Ky) )  \; ,
\label{eq:3.27}
\end{equation}
where $L = 2 \pi / K$ is their characteristic length-scale.  
The two above fields correspond to two typical situations: the shear
flow strongly enhances the large-scale diffusion coefficient (in the
shear direction), while the cellular flow (due to trapping) is
characterized by a weaker enhancement of the diffusivity.

Concerning the small-scale velocity component we consider a small
scale replica of the cellular flow (\ref{eq:3.27}), i.e.:
\begin{eqnarray}
{\bm u} = (u \sin(kx) \cos(ky), - u \cos(kx) \sin(ky)) \;,
\label{eq:3.1}
\end{eqnarray}
with characteristic length-scale given by $\ell = 2 \pi / k$ and
amplitude $u$.  In the absence of large-scale velocity fields and for
high Peclet numbers ($Pe = u \ell / D_0$), it is possible to
show~\cite{p85,S87} that this periodic array of small vortexes give rise
to an enhanced effective (small-scale) diffusivity $D^{S} \sim D_0
\sqrt{Pe}$. A precise estimation of $D^S$, which is indeed in good
agreement with the above expression, can be obtained by the numerical
solution of Eq.~(\ref{eq:4}), with ${\bm U}=0$. In particular, one
finds an isotropic eddy diffusivity induced by the small-scale
cellular flow $D^S_{ij} = D^S \delta_{ij}$. The numerical technique
used to solve Eq.~(\ref{eq:4}) is based on a standard implementation
of a fully de-aliased pseudospectral algorithm on a biperiodic square
lattices of size $L=2\pi$ with $512^2$ collocation points.

\begin{figure}[t!]
\centering
\includegraphics[scale=0.7]{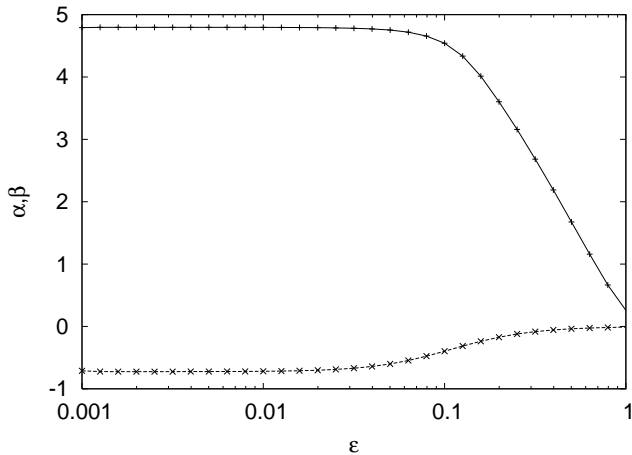}
\caption{$\alpha$ (solid line) e $\beta$ (dashed line) computed as a
function of the ratio $\varepsilon=U/u$.  Simulations have been
performed with a small scale cellular flow (\ref{eq:3.1}) with $u =1$
and $k=8$ (i.e. $ \ell = \pi /4$), the molecular diffusivity is $D_0 =
10^{-3}$ (corresponding to $Pe = 785$) and $\bm U=(U,0)$.}
\label{fig:correzioni2}
\end{figure}

Let us now discuss the effect on the eddy diffusivity tensor induced
by a weak large scale flow. 

In order to estimate the second-order correction and to assess the
limits of validity of truncating the perturbative expansion at second
order in $\varepsilon=U/u$, we proceeded as follows.  The full
auxiliary Eq.~(\ref{eq:4}) is solved for various choices of
$\varepsilon$. As already stated, the coefficients $\alpha$ and
$\beta$ defined in Eq.~(\ref{eq:20}) are independent (in the case of
small-scale isotropic flows) of the detailed spatio-temporal structure
of the large scale velocity, which is therefore taken, in a suitable
reference frame, as a constant flow in the $x$-direction ($\bm U =
(U,0)$). Notice that this is consistent with the assumption of  
complete scale separation, i.e. $\epsilon=0$, in which the small-scale
flow ``sees'' the large-scale velocity as a constant mean flow. 

The coefficients $\alpha$ and $\beta$, are then obtained by
normalizing with $U^2$ the eigen-values of $D_{ij}-D^S_{ij}$.  As
shown in Figure~\ref{fig:correzioni2} in a wide range of $\varepsilon$
values the coefficient $\alpha$ and $\beta$ do not depend on
$\varepsilon$, meaning that the second-order expansion holds. Of
course for larger values of $\varepsilon$ higher order corrections
must be taken into account. The results shown in
Fig.~\ref{fig:correzioni2} refer to the case of a small-scale cellular
flow with $u =1$, $ \ell = \pi /4 $, $D_0 =10^{-3}$ corresponding to a
large value of the Peclet number $Pe = u \ell /
D_0=785$. Qualitatively similar results have been obtained for other
sets of parameters.

\begin{figure}[t!]
\centering
\includegraphics[scale=0.7]{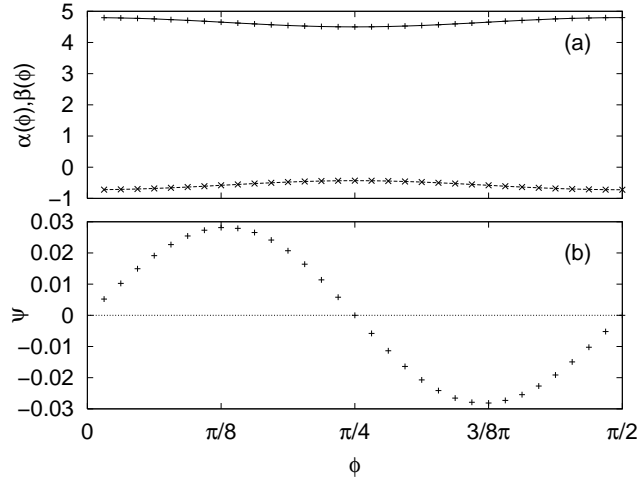}
\caption{(a) The coefficients $\alpha$ e $\beta$ as a function of
angle $\phi$ that defines the direction of the large-scale velocity
field.  (b) Angular discrepancy $\psi$ between the direction of the large-scale flow
and the eigen-direction of the eddy diffusivity tensor. Here the large
scale flow is chosen as $\bm U=U (\cos(\phi),\sin(\phi))$, where $U$
is fixed as $U=0.01 u$. The other parameters are as in
Fig.~\ref{fig:correzioni2} }
\label{fig:correzioni1}
\end{figure}

However, since the small-scale cellular flow is not exactly
isotropic. Therefore $\alpha$ and $\beta$ are expected to display a
weak dependence on the direction of the large-scale flow.  To test
this angular dependence, we repeated the measurements of $\alpha$ and
$\beta$ keeping fixed the ratio $\varepsilon=U/u$ (within the interval
of validity of the second order approximation) and varying the angle
$\phi$ between the direction of ${\bm U}$ and the $x$-axis of the
small scale cellular flow. As one can see in
Fig.~\ref{fig:correzioni1}a, the angular dependence is confirmed to be
very weak. Thus, for any practical use, the small scale cellular flow
can be considered almost isotropic.  We also measured the angular
discrepancy between the direction of ${\bm U}$ and the
eigen-directions of $\delta D^{(2)}_{ij}$ as a function of $\phi$ (see
Fig.~\ref{fig:correzioni1}b).  Also in this case the angular
dependence is indeed rather small.  Therefore, in the following we
shall ignore it and substitute $\alpha(\phi)$ and $\beta(\phi)$ with
their angular averages.

Once the correction to the small scale diffusivity tensor have been
parameterized through the two coefficients $\alpha$ and $\beta$, one can
reconstruct the spatial structure of the effective eddy diffusivity
for a generic large scale flow, within the second order approximation
by using Eq.~(\ref{eq:20}). 

\begin{figure}[t!]
\centering
\includegraphics[scale=0.7]{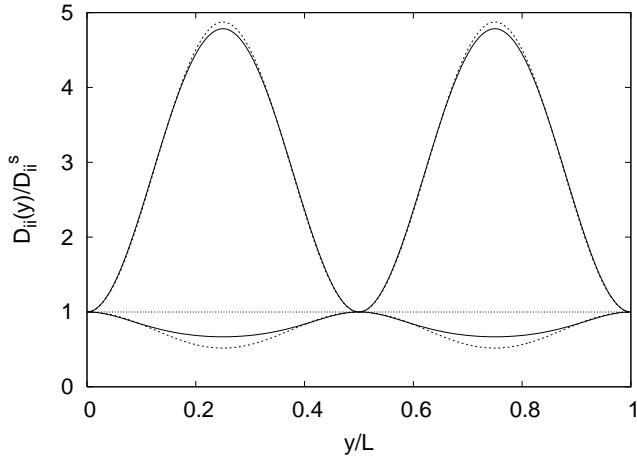}
\caption{The eddy-diffusivity $D_{ii}(y)$ normalized with the
small-scale diffusivity $D_{ii}^S$.  The second-order approximation
(dashed line) for $D_{xx}(y)$ (above $1$) and $D_{yy}(y)$ (below $1$)
is in almost perfect agreement with the exact multiple-scale solution
(solid line).  Data refer to the case of a small-scale cellular flow
(\ref{eq:3.1}) superimposed to a large-scale shear (\ref{eq:3.26}),
with parameters values $U=0.1$, $L=2 \pi$, $u=1$, $\ell =L/8$, and
$D_0 = 10^{-3}$. The scale separation is $\epsilon=1/8$ (see text for
a discussion about this point).}
\label{fig:deff}
\end{figure}

In Figure~\ref{fig:deff} we show the eddy-diffusivity $D_{ij}(y)$
resulting from the small-scale cellular flow (\ref{eq:3.1})
superimposed to a large-scale shear (\ref{eq:3.26}) . We compare the
exact solution of Eq.~(\ref{eq:4}), $D_{ij}(y)$, with its second-order
approximation for $\varepsilon=0.1$.  As one can see in
Fig.~\ref{fig:deff}, though we are at the border of the validity
interval of the perturbation theory (see Fig.~\ref{fig:correzioni2}),
the second-order approximation recovers quite well the exact
multiple-scale solution, both in the parallel and perpendicular
direction with respect to the large-scale shear.  It should be
remarked that the improvement brought by the second-order
approximation with respect to the zeroth-order one (the constant
dotted line in Fig.~\ref{fig:deff}) is impressive.

It is worth noticing that the corrections induced by the large scale
are not only non-homogeneous but also non-isotropic. In particular,
the diffusion is enhanced in the longitudinal direction (being
$\alpha>0$) while it is decreased in the transverse direction
($\beta<0$). As we shall see in the next section this anisotropy is
crucial to recover the pre-asymptotic features of scalar transport.

\section{Asymptotic and non-asymptotic scalar transport}
\label{sec:results}
Up to now the discussion essentially focused on the formalism and the
methods to obtain the effective meso-scale description, it is now
natural to wonder about the practical use of this approach.  As first,
to test the consistency of the approach, we check whether the model
recovers the asymptotic behavior of the scalar field at very large
scales $\mathcal{L} \gg L$ and very long times $\mathcal{T}\gg
T$. Then we shall show to what extent Eq.~(\ref{eq:2}) is a good model
for the meso-scale transport, i.e. in the interval $\ell \ll r \leq
L$. In particular, we shall show that, only properly taking into
account the effects of the large-scale flow on the diffusivity tensor
(i.e. at the second order Eq.~(\ref{eq:11})), the meso-scale evolution
of the scalar field can be correctly predicted.

\subsection{Macro-dynamics of scalar transport}
\label{sec:macro}
As it is well known (see Ref.~\cite{MK99} for a review on the
subject), in the asymptotics of large scales and long times, scalar
dynamics reduce to an effective diffusion equation :
\begin{equation}
\partial_{\mathcal{T}} \theta_{\mathcal{L}} = 
D^{\mathcal{L}}_{ij} \nabla_i \nabla_j \theta_{\mathcal{L}} \;,
\label{eq:macro}
\end{equation}
where $\theta_{\mathcal{L}}$ is the scalar field averaged over volumes
of size $L$. Following the strategy devised in Ref.~\cite{MMV05} (see
also the Appendix), we notice that the very large scale diffusivity
tensor $D^{\mathcal{L}}_{ij}$ can be obtained in two ways. The first
is the obvious implementation of the multiple-scale analysis directly
to Eq.~(\ref{eq:1}). On the other hand one may proceed by using two
successive homogenization steps. In the first one wipes out the
small-scale details of the velocity field, this is equivalent to
obtaining Eq.~(\ref{eq:2}). The second step consists in applying the
multiple scale technique to the meso-scale equation (\ref{eq:2}).  In
principle if more than two scales are present the method can be
iterated (see Ref.~\cite{MMV05} for a detailed discussion of this
point).  Both procedures lead to an effective diffusive equation
(\ref{eq:macro}) but with two (a priori) different eddy diffusivity
tensors.

\begin{figure}[t!]
\centering
\includegraphics[scale=0.7]{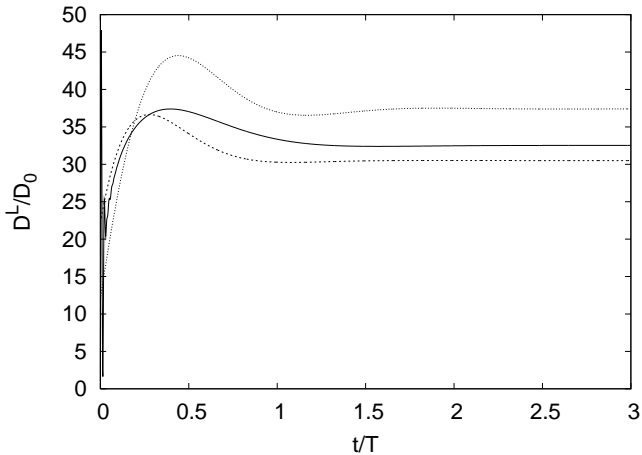}
\caption{Time evolution of the asymptotic eddy-diffusivity, computed
via Eqs.~(\ref{eq:A3}) and~(\ref{eq:A1}), up to convergence to its
constant value (non-dimensional units).  The velocity field is given
by the superimposition of a large-scale cellular flows
(\ref{eq:3.27}) and a small-scale one (\ref{eq:3.27}), with parameters
values $U=0.1$, $L=2 \pi$, $u=1$, $\ell =L/8$, and $D_0 = 10^{-3}$.
The time evolution of the asymptotic eddy-diffusivity $D^{\mathcal L}$
(solid line) is well approximated by meso-scale modeling~(\ref{eq:2})
in which the second-order correction induced by the large scale flow
are retained (dashed line), while the zeroth-order approximation
(dotted line) does not match the actual value.  Time is normalized
with the large scale advecting time scale $T=L/U$.  }
\label{fig:asym}
\end{figure}

The former method gives the exact value of the eddy-diffusivity tensor
$D^{{\mathcal L},ex}$ (see Eq.~(\ref{eq:A3})), but requires the detailed
knowledge of the velocity field at both large and small scales, which
is a far too unrealistic situation in applications. The second
procedure (see Eq.~\ref{eq:A1}) is less accurate but is based on the sole
large-scale velocity, $\bm{U}$, and the eddy-diffusivity $D_{ij}({\bm
X},T)$ that parameterizes the small-scale flow and can be obtained 
with the procedure described in the previous section.

In order to test whether the meso-scale model based on the
perturbative expansion (\ref{eq:11})  is
able to recover the asymptotic behavior of the scalar field, we
compare the exact value $D^{{\mathcal L},ex}$ obtained by a direct
homogenization of Eq.~(\ref{eq:1}), with the approximations
$D^{{\mathcal L},M0}$ and $D^{{\mathcal L},M2}$.  These latter are
obtained by homogenization of meso-scale model (\ref{eq:2}) where the
zeroth-order approximation $D^{(M0)}_{ij}=D^S_{ij}$ and the
second-order approximation $D^{(M2)}_{ij}=D^S_{ij}+\delta
D^{(2)}_{ij}$ are used for the eddy-diffusivity (see
Eqs.~(\ref{eq:20})).

Results are summarized in Fig.~\ref{fig:asym}, where we show the time
evolution of the asymptotic eddy-diffusivity up to its convergence to
a constant value in the case of the large-scale
cellular flow (\ref{eq:3.27}). Notice that by using the second-order
approximation one obtains a $6\%$ discrepancy from the exact value
with respect to a $15\%$ for the zeroth-order approximation. In
Table~\ref{table} one may directly compare the different
approximations for the case of large-scale cellular and shear flow.

\begin{table}[t!]
\begin{center}
\begin{tabular}{|c|c|c|c|c|c|} 
\hline
Type &  $\ell /L$ & $U/u$ & $D^{{\mathcal L},ex}$ & $D^{{\mathcal L},M2}$ & $D^{{\mathcal L},M0}$ \\
\hline
S & $\frac{1}{8}$ &  $0.1$ & $D_{xx} = 0.486$ & $0.520$ & $0.429$ \\
& & & $D_{yy} = 9.53 \cdot 10^{-3} $ &  $ 8.64 \cdot 10^{-3} $ & $ 1.20 \cdot 10^{-2} $ \\
\hline
C & $\frac{1}{8}$ & $0.1$ & $D_{xx} = D_{yy} = 3.25 \cdot 10^{-2} $ & $ 3.05 \cdot 10^{-2} $ & $ 3.74 \cdot 10^{-2} $ \\
\hline
\end{tabular}
\caption{Asymptotic eddy-diffusivity resulting from the 
small-scale cellular flow~(\ref{eq:3.1}) ($u=1.0$, $\ell= \pi/4$) and large-scale (S)
shear~(\ref{eq:3.26})  
or (C) cellular flow~(\ref{eq:3.27}) ($U=0.1$, $L= 2 \pi$). 
In all cases we used molecular diffusivity $D_0=10^{-3}$. }
\label{table}
\end{center}
\end{table}

In all the investigated cases taking into account the effect of the
large scales, even if at the lowest nontrivial order (i.e. the second
one) allows for an improvement of at least a factor $2$
in the relative error. Though not astonishing this goes in the correct
direction and confirms the consistency of the approach. We conclude by
noticing that here we are comparing ``global'' quantities while we
expect a better performance of the meso-scale model for local properties at
the intermediate scales, for which the model itself has been developed for.

\subsection{Meso-scale dynamics of scalar transport}
\label{sec:meso}
In many applications the asymptotic properties are less
important than the meso-scale ones.  For instance, let us consider an
initially localized concentration field (as, e.g.,
a pollutant released in a given region). More than being interested in
the time-scales necessary for it to distribute uniformly in the whole
domain, one is interested in predicting the spatial patterns and evolution of
the pollutant concentration at intermediate time and scales (this is
dramatically important if, e.g., the pollutant is a toxic substance).

In this perspective it is interesting to see if the model (\ref{eq:2})
has a ``predictive'' character for the scalar dynamics at these
intermediate scales and finite times. In order to test such a
possibility we devised the following strategy. We consider a scalar
concentration initially localized at the small scales.  Then we follow
its evolution in a square domain with periodic boundary conditions
according to:
\begin{description}
\item{i)}   the exact dynamics of the scalar field $\theta_{(T)}$ 
            given by Eq.~(\ref{eq:1});
\item{ii)}  the model dynamics given by Eq.~(\ref{eq:2}) with 
            $D_{ij}=D^{(M0)}_{ij}=D^S_{ij}$, in the
            following this is called the $\theta_{(M0)}$ field;
\item{iii)} the model dynamics given by Eq.~(\ref{eq:2}) with the
            refined approximation for the diffusivity tensor
            $D_{ij}=D^{(M2)}_{ij} = D^S_{ij}+\delta D^{(2)}_{ij}$ 
            (see Eqs.~(\ref{eq:20})), in the following we denote as
            $\theta_{(M2)}$ the resulting field.
\end{description}
Finally the fields evolving with the three above dynamics are
compared. The comparison is done at different levels. We looked both
at the evolution of local and global quantities. In particular, we
compare the evolution of the variance of the three fields, i.e.
\begin{equation}
\sigma^2_{(\alpha)}(t)=\langle \theta^2_{(\alpha)}(\bm x,t)\rangle-
\langle \theta_{(\alpha)}(\bm x,t)\rangle^2\,,\qquad \alpha=T,M0,M2\,,
\label{eq:energy}
\end{equation}
where the average is performed over the spatial domain. The behavior
of $\sigma_{(\alpha)}^2(t)$ describes the mean decay of the scalar
fluctuations due to the joint effects of molecular dissipation and
advection. Natural indicators to characterize  the degree of 
spatial ``similarity'' between the different fields are:
\begin{eqnarray}
E_{(M0)}(t) &=& \langle (\theta_{(T)}(\bm x,t)-\theta_{(M0)}(\bm
x,t))^2\rangle/\sigma_{(T)}^2(t) \nonumber \\ E_{(M2)}(t) &=& \langle
(\theta_{(T)}(\bm x,t)-\theta_{(M2)}(\bm
x,t))^2\rangle/\sigma_{(T)}^2(t) \label{eq:error}\,.
\end{eqnarray}
With the normalization by $\sigma_{(T)}^2(t)$, $E_{(M0)}(t)$ and
$E_{(M0)}(t)$ provide a measure of the relative distance between true
and model fields.  It is worth noticing that, since $\theta_{(M0)}$
and $\theta_{(M2)}$ models the ``true'' evolution only at scales
larger than $\ell$, in principle one should measure the errors
(\ref{eq:error}) with a filtering out these small scales from
$\theta_{(T)}$. Here, to avoid the arbitrariness of the choice of
filtering procedure, we did used the unfiltered field
$\theta_{(T)}$. This introduces an additional error caused by the
fast-decorrelating small-scales features, which nevertheless becomes
negligible as soon as the concentrations $\theta_{(\alpha)}$ spreads
over scales significantly larger than $\ell$.

\begin{figure}[t!]
\centering \includegraphics[scale=0.54]{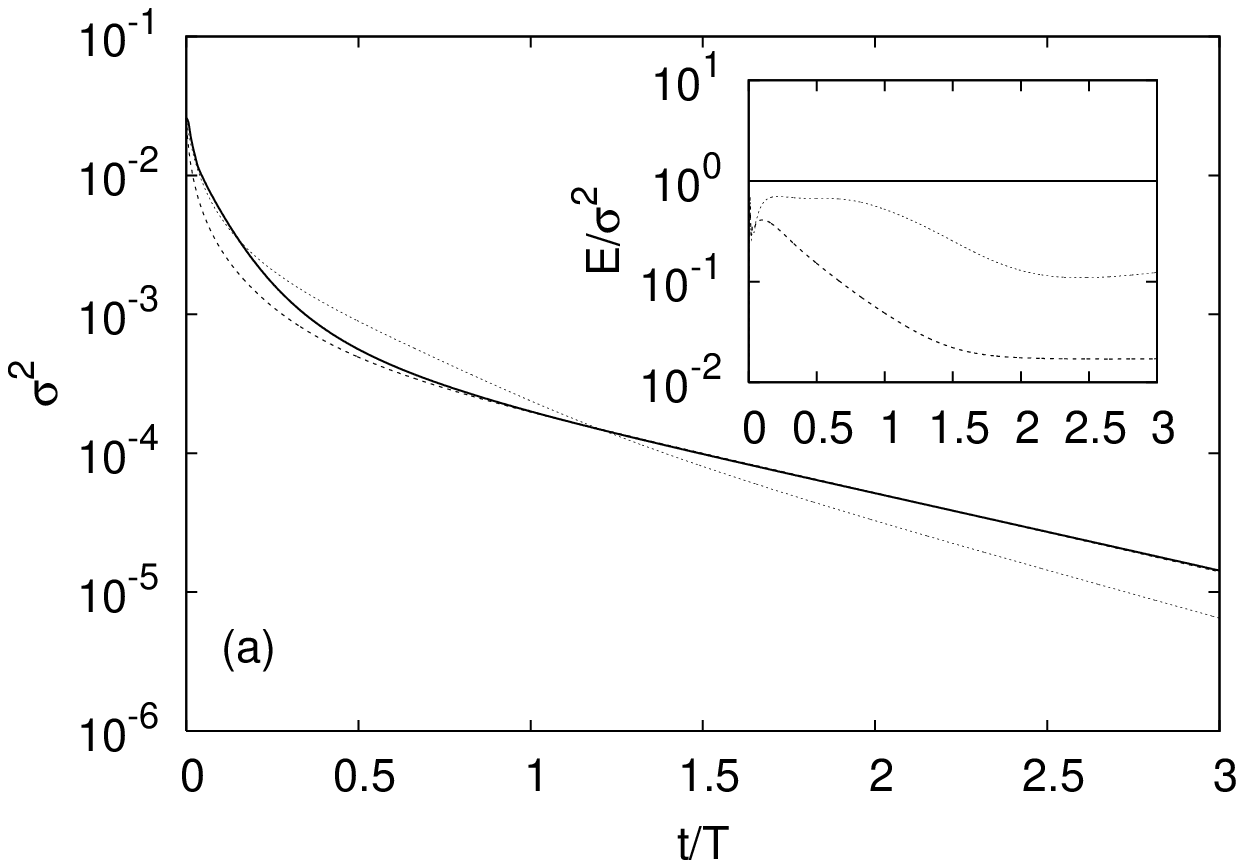}\hfill
\includegraphics[scale=0.54]{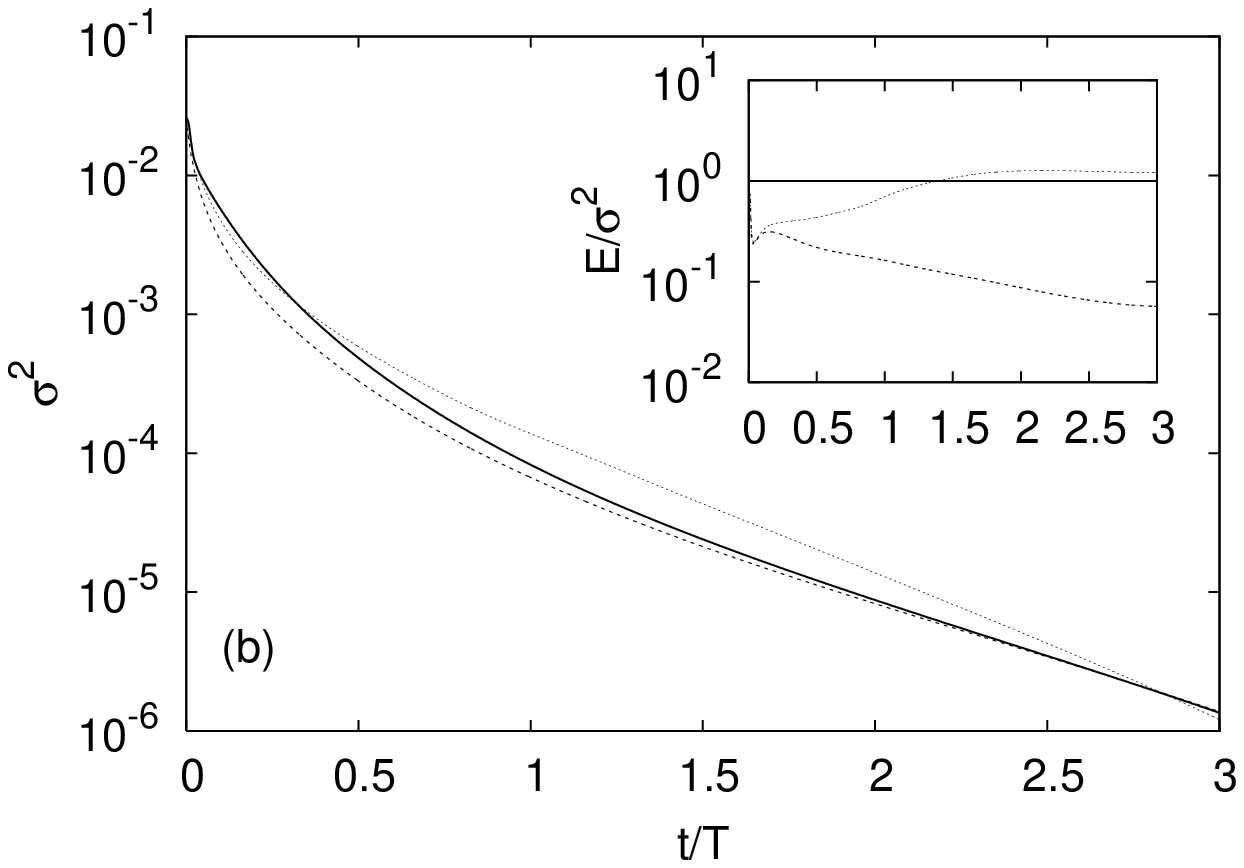}
\caption{(a) Temporal decay of the fluctuations, $\sigma^2_{(T)}$
(bold line) $\sigma^2_{(M0)}$ (dotted line) and $\sigma^2_{(M2)}$
(dashed line), in a large scale shear and a small scale cellular flow.
The inset shows the normalized error $E_{(M0)}(t)$ (dotted line) and
$E_{(M2)}(t)$ (dashed line).  (b) The same for the case of large scale
and small scale cellular flows. Parameters as in table~\ref{table}.
Time is normalized with the large scale advecting time scale $T=L/U$.}
\label{fig:decay}
\end{figure}

On the numerical side, the three fields are identically initialized as
Gaussian distributions with width $\sim O(\ell)$ and centered in the
same point of the domain. The robustness of the results we are going
to present has been tested by repeating the computation with different
initial locations. In the following we present the results for both 
the large-scale shear (\ref{eq:3.26}) and cellular flow (\ref{eq:3.27}).

In Figs.~\ref{fig:decay}a,b we show the time evolution of the above
defined indicators. As one can see after a short transient the
second-order meso-scale modeling recovers the actual decay rate, while
the zeroth-order approximation does not. Moreover, as evidenced in the
insets, the relative errors between the true evolving field and its
two models  M0 and M2 is such that, while the zeroth-order
approximation rapidly goes toward $100\%$ error, the second-order one
remains below $15\%$ during the whole evolution.

\begin{figure}[t!]
\centering
\includegraphics[draft=false,scale=.55,clip=true]{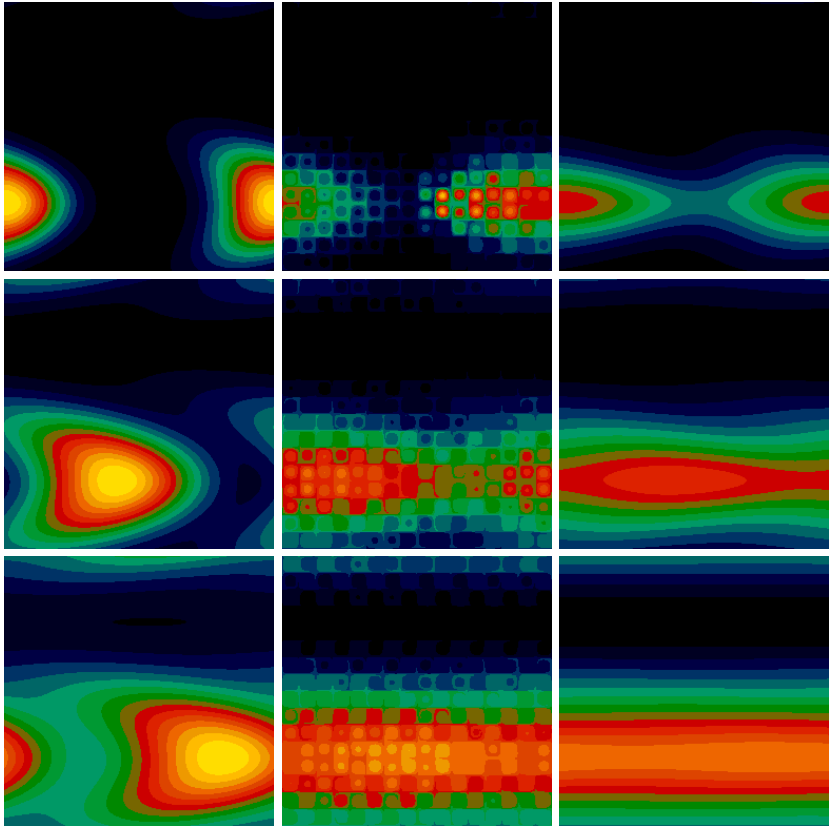}
\caption{Snapshots of $\theta_{(M0)}$ (left column),
$\theta_{(T)}$ (middle column), and $\theta_{(M2)}$ (right column) at
three different time, from top to bottom $t\approx 0.5, 1.0, 1.5 T$
where $T=L/U$ is the large scale advection time. The large-scale
velocity field is the shear flow (\ref{eq:3.26}). At $t=0$
$\theta_{\alpha}$'s were initialized as a Gaussian centered in
$(9/16L,7/16L)$ with a width $2\sigma\approx \ell$.
Simulation parameters are summarized in table~\ref{table}.
}
\label{fig:snapS}
\end{figure}

\begin{figure}[t!]
\centering
\includegraphics[draft=false,scale=.55,clip=true]{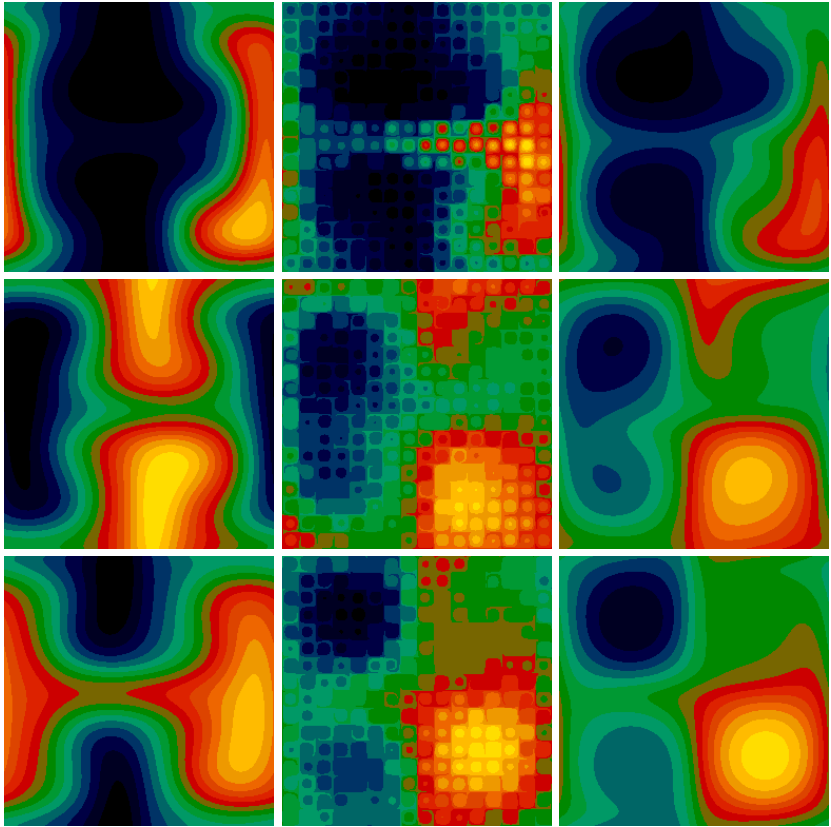}
\caption{The same as in Fig.~{\ref{fig:snapS}} for the case of large-scale cellular flow.
Simulation parameters are summarized in table~\ref{table}.
}
\label{fig:snapG}
\end{figure}

The improvement brought by the second-order corrections is even more
striking if one looks directly to the snapshots of the concentrations
fields (Figs.~\ref{fig:snapS},~\ref{fig:snapG}). As one can see the
spatial patterns of the M0 approximation rapidly decorrelate with
those of the true field, that are actually well described by the M2
approximation.  We remark crucial for the fidelity of the second-order
approximation is that the corrections to the small-scale diffusivity
tensor retains relevant information of the anisotropic and
inhomogeneous diffusive behavior induced by the presence of the
large-scale flow, namely the enhancement of diffusion in the direction
on $\bm{U}$ and its reduction in the transverse direction.  This is
particularly evident in the case of large-scale shear shown in
Fig.~\ref{fig:snapS}, where one can see that the blob M0 spreads too
quickly in the $y$-direction and too slowly in the $x$-direction with
respect to the true field. On the other hand, since $D^{(M2)}_{xx}
>D^{(M0)}_{xx}$ and $D^{(M2)}_{yy} <D^{(M0)}_{yy}$ these features are
captured by the M2 approximation.  This is even clearer in
Fig.~\ref{fig:snapG}, where one can see that, differently from the M2
model, the trapping of the concentration in the large-scale cells is
completely missed in the M0 approximation.

\section{Conclusions}
\label{sec:conclu}

We have studied both analytically and numerically the effects of a
weak advecting velocity field at large scale on the meso-scale
modeling for the transport of passive scalars. By means of
multiple-scale methods we perturbatively computed the dependence of
(pre-asymptotic) eddy diffusion tensor $D_{ij}$ on the large-scale
velocity field. The corrections to $D$ are non-homogeneous 
and non-isotropic. In particular we find an enhancement (reduction) 
in the longitudinal (transversal) direction of the large-scale field.
The perturbative approach proposed here allows to develop meso-scale
models retaining (at least at second order accuracy) these effects, 
which are shown to be crucial for properly describe the transport dynamics.  

We conclude by noticing that the new findings obtained by our approach
seem very promising for future applications to the numerical
investigation of large-scale transport (asymptotic and pre-asymptotic)
both in the atmosphere and in the ocean. Indeed the present results
together with those obtained in Ref.~\cite{MMV05} cover two opposite
limit of transport for which explicit expressions for the effective
parameters are available. By interpolations, one may hope to obtain
the form of these effective coefficients under general conditions.

\section*{Acknowledgments}
We acknowledge partial support from MIUR Cofin2003 ``Sistemi Complessi
e Sistemi a Molti Corpi'', and EU under the contract
HPRN-CT-2002-00300.  MC acknowledges the Max Planck Institute for the
Physics of Complex Systems for computational resources.
                                                                               

\appendix
\section{Expressions for the asymptotic diffusivity}

There are two, in principle non equivalent, ways to obtain
the large-scale equation (\ref{eq:macro}). The first one is to 
apply the homogenization technique to the exact equation
(\ref{eq:1}), while the
second possibility is to start  from the meso-scale model equation 
(\ref{eq:2}). 
Following the first approach
the (exact) value of the eddy-diffusivity tensor, $D^{{\mathcal L},ex}$, 
depends on both the  
molecular diffusivity and the advection by the total velocity field
${\bm v} = {\bm U} + {\bm u}$: 
\begin{equation}
D^{{\mathcal L},ex}_{ij} = D_0 \delta_{ij}
- \frac{\langle v_i \chi_j \rangle + \langle v_j \chi_i \rangle}{2} \;. 
\label{eq:A3}
\end{equation}
The auxiliary field ${\bm \chi}$ 
is the solution of the following equation
\begin{equation}
\partial_t {\bm \chi} + ({\bm v} \cdot {\bm \partial} ){\bm \chi} -
D_0 \partial^2 {\bm \chi} = - {\bm v} \;.
\label{eq:3.15}
\end{equation}
With the second way the asymptotic
eddy-diffusivity tensor $D^{\mathcal{L}}$  results from the
combined effects of the advection given by the large-scale flow
$\bm{U}({\bm X},T)$ and the effective diffusion at scale $\ell$ 
that depends on space and time through $D_{ij}({\bm X},T)$ 
(see \cite{M97} for a detailed derivation),
\begin{equation} 
D^{\mathcal{L}}_{ij} = 
- \frac{\langle U_i \chi_j \rangle + \langle U_j \chi_i \rangle}{2} 
+ \frac{\langle D_{ik} \partial_k \chi_j \rangle +
   \langle D_{jk} \partial_k \chi_i \rangle}{2}
+ \frac{\langle D_{ij} \rangle + \langle D_{ji} \rangle}{2} \,.
\label{eq:A1}
\end{equation}
Here the vector field ${\bm \chi}$ is obtained by the auxiliary
equation
\begin{equation}
\partial_t \chi_k + ({\bm U} \cdot {\bm \partial} ) \chi_k 
- \partial_i (D_{ij} \partial_j \chi_k) = - U_k + \partial_i D_{ik} \;.
\label{eq:3.13}
\end{equation}
It should be noted that even though the first procedure gives the
exact value of the eddy-diffusivity tensor $D^{{\mathcal L},ex}$, it
requires the detailed knowledge of the velocity field at both large
and small scales. However, the expression obtained from
Eq.~(\ref{eq:A1}) (which, in general, does not coincide with
$D^{{\mathcal L},ex}$) has the advantage of being based solely on the
large-scale velocity, $\bm{U}$, with the effects of the small-scale
flow  included in the eddy-diffusivity $D_{ij}({\bm X},T)$.



\begin{thebibliography}{99}

\bibitem{MRAK02}
J.S.~Martins  J.B.~Rundle M.~Anghel, W.~Klein,
Phys. Rev. E {\bf 65}, (2002) 056117. 

\bibitem{T97}
D.L.~Turcotte, 
{\it Fractals and Chaos in Geology and Geophysics},
(2nd ed., Cambridge Univ. Press, New York, 1997).

\bibitem{MFPKV04}
T. Murtola, E. Falck, M. Patra, M. Karttunen, and I. Vattulainen,
J. Chem. Phys. {\bf 121}, (2004) 9156.

\bibitem{O89}
J.M.~Ottino, {\it The Kinematics of Mixing: Stretching, Chaos and Transport},
(Cambridge Univ. Press, New York, 1989).

\bibitem{LES}O.~Metais and J.~Ferziger {\it New Tools in Turbulence
Modelling: Les Houches School} (Springer-Verlag, Berlin 1996)

\bibitem{MK99}
A.~J. Majda and P.~R. Kramer.   
Phys. Rep. {\bf 314} (1999) 237.

\bibitem{S88} 
R.B.~Stull, 
{\it  An Introduction to Boundary Layer Meteorology}, 
(Kluwer Academic Publications, 1988).

\bibitem{BLP78}
A.\ Bensoussan, J.-L.\ Lions and G.\ Papanicolaou,
{\it Asymptotic Analysis for Periodic Structures}
(North-Holland, Amsterdam, 1978).

\bibitem{Piretal}
D.~Mc Laughlin, G.C.~Papanicolaou  and  O.~Pironneau,
SIAM Journal of Appl. Math. {\bf 45} (1985) 780.

\bibitem{BGW89} 
R.~N. Bhattacharya, V.~K. Gupta, and H.~F. Walker, 
SIAM J. Appl. Math. {\bf 49} (1989) 86.

\bibitem{BCVV95}
L.~Biferale, A.~Crisanti, M.~Vergassola and A.~Vulpiani, 
Phys. Fluids {\bf 7} (1995) 2725 .

\bibitem{VA97} 
M. Vergassola and M. Avellaneda, Physica D {\bf 106} (1997) 148.

\bibitem{PK02}
G. A. Pavliotis and P.R. Kramer, ``Homogenized transport by a
spatiotemporal mean flow with small-scale periodic fluctuations'',
in proc. of the IV international conference on dynamical systems and
differential equations, pp. 1-8, 
May 24 - 27, 2002, Wilmington, NC, USA.

\bibitem{M03}
R. Mauri, Phys. Rev. E {\bf 68} (2003) 066306.

\bibitem{MY75} 
A.S.~Monin, and A.~M. Yaglom, 
{\it Statistical fluid mechanics}
(vol. II, MIT, Press, Cambridge, 1975).

\bibitem{M97} 
A.~Mazzino, Phys. Rev. E {\bf 56} (1997) 5500.

\bibitem{MMV05} A. Mazzino, S. Musacchio and A. Vulpiani, Phys. Rev. E
{\bf 71} (2005) 011113.

\bibitem{L96}
E.N. Lorenz, ``Predictability - a problem partly solved'' {\it Proc. Seminar on
predictability,} Reading, U.K., European Centre for Medium-Range Weather
Forecast, (1996) 1-18.

\bibitem{BCFV02} 
G.~Boffetta, M.~Cencini, M.~Falcioni and A.~Vulpiani, 
Phys. Rep. {\bf 356} (2002) 367.

\bibitem{HS00}
P.~Haynes and E.~Shuckburgh,
J. Geophys. Res, {\bf 105} (2000) 22777.

\bibitem{MSJH05}
J.~Marshall, E.~Shuckburgh, H.~Jones and C.~Hill, 
(submitted to J. Phys. Oceanog.)

\bibitem{MS61}
L. Meshalkin and Ya G. Sinai, J. Appl. Math. Mech. {\bf 25}, (1961) 1700.

\bibitem{sg88}
T.H. Solomon and J.P. Gollub Phys. Rev. A {\bf 38} (1988) 6280.

\bibitem{p85}
B. Y. Pomeau, C.R. Acad. Sci. Paris {\bf 301} (1985) 1323.

\bibitem{S87}
B.I. Shraiman, Phys. Rev. A {\bf 36} (1987) 261. 

\end{thebibliography}
\end{document}